\documentclass[twocolumn,twoside,letterpaper,11pt]{article}
\usepackage{graphicx,xrrc,natbib}
\usepackage{times}

\def\la{\mathrel{\hbox{\rlap{\hbox{\lower4pt\hbox{$\sim$}}}\hbox{$<$}}}}
\def\ga{\mathrel{\hbox{\rlap{\hbox{\lower4pt\hbox{$\sim$}}}\hbox{$>$}}}}

\begin{document}

\title{MERGERS AND NON-THERMAL PROCESSES IN CLUSTERS OF GALAXIES}

\author{    C. L. Sarazin                                } 
\institute{ University of Virginia                       } 
\address{   P.O. Box 3818, Charlottesville, VA 22903-0818, U.S.A.        } 
\email{     sarazin@virginia.edu }

\maketitle

\abstract{Clusters of galaxies generally form by the gravitational merger
of smaller clusters and groups.
Major cluster mergers are the most energetic events
in the Universe since the Big Bang.
The basic properties of cluster
mergers and their effects will be discussed.
Mergers drive shocks in
the intracluster gas, and these shocks heat the intracluster gas, and
should also accelerate nonthermal relativistic particles.  Mergers also
produce distinctive features in the X-ray images of clusters, including
``cold fronts'' and cool trails.
Chandra and XMM/Newton observations of
the X-ray signatures of mergers will be discussed.
X-ray observations of shocks and cold fronts can be used to determine the
geometry and kinematics of the merger.
As a result of particle acceleration in shocks and turbulent
acceleration following mergers, clusters of galaxies should contain very
large populations of relativistic electrons and ions.  Observations and
models for the radio, extreme ultraviolet, hard X-ray, and gamma-ray
emission from nonthermal particles accelerated in these shocks will also
be described.}

\section{Introduction}
\label{sec:8.1_sarazin_intro}

\begin{figure}
\begin{center}
\includegraphics[width=\columnwidth]{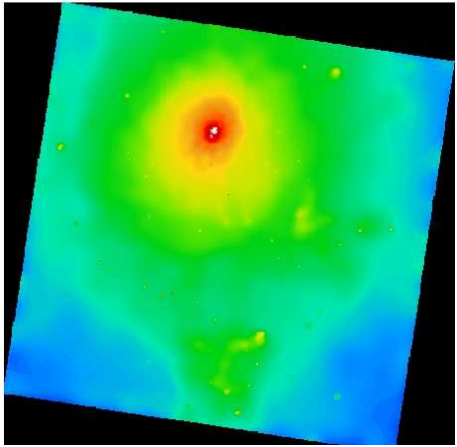}
\caption{\small The Chandra X-ray image of the merging cluster Abell~85
(Kempner, Sarazin, \& Ricker 2002).
Two subclusters to the south and southwest are merging with the main
cluster.}
    \label{fig:8.1_sarazin_a85}
  \end{center}
\end{figure}

\begin{figure}
\begin{center}
\includegraphics[width=\columnwidth]{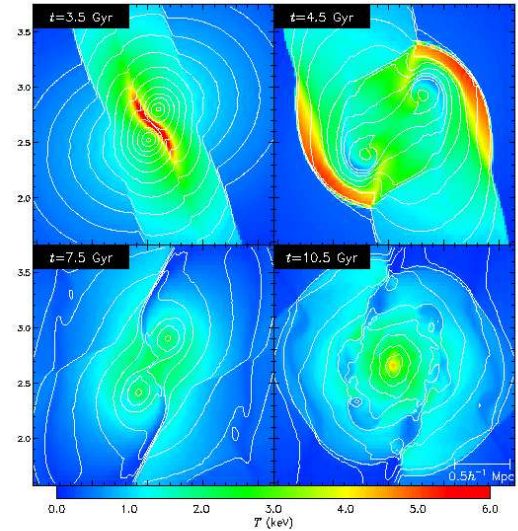}
\caption{\small
A hydrodynamical simulation of a cluster merger from
Ricker \& Sarazin (2001).
The hot (red) regions are merger shocks.}
    \label{fig:8.1_sarazin_sim}
  \end{center}
\end{figure}

Clusters of galaxies form hierarchically by the merger of smaller
groups and clusters.
Major cluster mergers are the most energetic events in the Universe
since the Big Bang.
In these mergers, the subclusters collide at velocities of
$\sim$2000 km/s,
releasing gravitational binding energies of as much as $\ga$$10^{64}$
ergs.
Figure~\ref{fig:8.1_sarazin_a85} shows the Chandra image of the merging cluster
Abell~85, which has two subclusters merging with the main
cluster
(Kempner, Sarazin, \& Ricker 2002).
The relative motions in mergers are moderately supersonic,
and shocks are driven into the intracluster medium.
Figure~\ref{fig:8.1_sarazin_sim} shows a numerical hydrodynamical
simulation of the cluster merger, showing the hot shocks propagating
through the merging clusters
(Ricker \& Sarazin 2001).
In major mergers, these hydrodynamical shocks dissipate energies of
$\sim 3 \times 10^{63}$ ergs; such shocks are the major heating
source for the X-ray emitting intracluster medium.
Merger shocks heat and compress the X-ray emitting intracluster
gas, and increase its entropy.
We also expect that particle acceleration by these shocks will produce
nonthermal electrons and ions, and these can produce synchrotron
radio, inverse Compton (IC) EUV and hard X-ray, and gamma-ray emission.

\section{Thermal effects of mergers}
\label{sec:8.1_sarazin_thermal}

Mergers heat and compress the intracluster medium.
Shocks associated with mergers also increase the entropy of the
gas.
Mergers can help to mix the intracluster gas, possibly removing abundance
gradients.
Mergers appear to disrupt the cooling cores found in many clusters;
there is an anticorrelation between cooling core clusters and clusters
with evidence for strong ongoing mergers
(e.g., Buote \& Tsai 1996).
The specific mechanism by which cooling cores are disrupted is not
completely understood at this time
(e.g., Ricker \& Sarazin 2001).

\begin{figure}
\begin{center}
\includegraphics[width=\columnwidth]{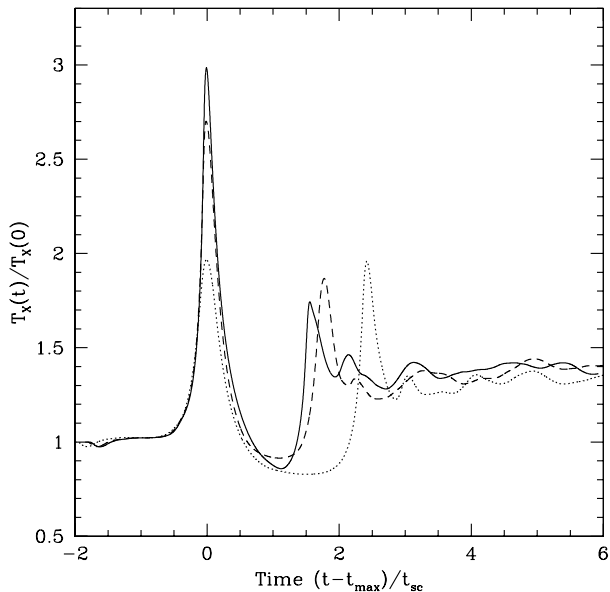}
\caption{\small The X-ray emission-averaged temperature in a
pair of equal mass clusters undergoing a merger
(Ricker \& Sarazin 2001;
Randall et al.\ 2002).
During a merger, the temperature can undergo a transcient boost by up
to a factor of three and the X-ray luminosity by up to a factor of ten.}
    \label{fig:8.1_sarazin_tboost}
  \end{center}
\end{figure}

The heating and compression associated with mergers can produce a
large, temporary increase in
the X-ray luminosity (up to a factor of $\sim$10) and
the X-ray temperature (up to a factor of $\sim$3) of the merging clusters
(Figure~\ref{fig:8.1_sarazin_tboost}; Ricker \& Sarazin 2001;
Randall, Sarazin, \& Ricker 2002).
Very luminous hot clusters are very rare objects in the Universe.
Although major mergers are also rare events, merger boosts can cause
mergers to strongly affect the statistics of the most luminous, hottest
clusters.
Simulations predict that many of the most luminous, hottest clusters
are actually merging systems, with lower total masses than would be
inferred from their X-ray luminosities and temperatures
(Randall et al.\ 2002).
Since the most massive clusters give the greatest leverage in determining
the cosmological parameters
$\Omega_M$ and $\sigma_8$, these values can be biased by merger boosts.
Recent weak lensing studies appear to have confirmed these large merger
boosts
(e.g., Smith et al.\ 2003).

Cluster mergers can also boost the Sunyaev-Zeldovich effect and
particularly the cross-section for a cluster to have strong
lensing
(Randall, Sarazin, \& Ricker 2004; Torri et al.\ 2004).

\section{Cold fronts and cool trails}
\label{sec:8.1_sarazin_cold}

\begin{figure}
\begin{center}
\includegraphics[width=\columnwidth]{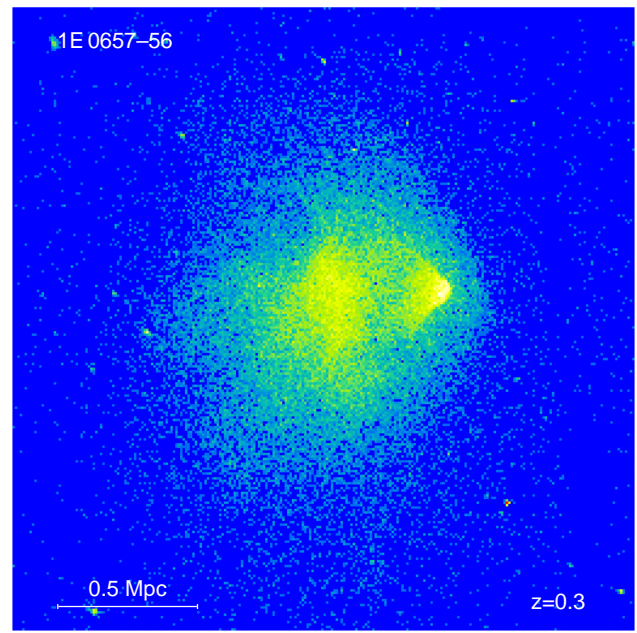}
\caption{\small The Chandra X-ray image of the ``bullet'' cluster 1E0657-56
(Markevitch et al.\ 2004).
The cluster shows a prominent cold front with a merger bow shock
ahead (to the right) of it.
(Figure kindly provided by Maxim Markevitch.)}
    \label{fig:8.1_sarazin_1e}
  \end{center}
\end{figure}

One of the most dramatic results on clusters of galaxies to come from
the Chandra X-ray observatory was the discovery of sharp surface brightness
discontinuities in the images of merging clusters
(Figures~\ref{fig:8.1_sarazin_a85} \& \ref{fig:8.1_sarazin_1e}).
These were first seen in Abell 2142
(Markevitch et al.\ 2000)
and Abell 3667
(Vikhlinin, Markevitch, \& Murray 2001b).
Initially, one might have suspected these features were merger shocks,
but X-ray spectral measurements showed that the dense, X-ray bright
``post-shock'' gas was cooler, had lower entropy, and was at the same
pressure as the lower density ``pre-shock'' gas.
This would be impossible for a shock.
Instead, these ``cold fronts'' are apparently contact discontinuities
between
gas which was in the cool core of one of the merging subclusters and
surrounding shocked intracluster gas
(Vikhlinin et al.\ 2001b).
The cool cores are plowing rapidly through the shocked cluster gas,
and ram pressure sweeps back the gas at the front edge of the cold front.
In a few cases (e.g., 1E0657-56; Figure~\ref{fig:8.1_sarazin_1e};
Markevitch et al.\ 2004), bow shocks are seen ahead of the cold fronts.

Cold fronts and merger shocks provide a number of classical hydrodynamical
diagnostics which can be used to determine the kinematics of the merger
(Vikhlinin et al.\ 2001b;
Sarazin 2002).
Most of these diagnostics give the merger Mach number $\cal M$.
The standard Rankine-Hugoniot shock jump conditions can be applied to
merger shocks; for example, the pressure discontinuity is
\begin{equation}
\frac{P_2}{P_1} = \frac{ 2 \gamma}{\gamma + 1} {\cal M}^2 -
\frac{\gamma - 1}{\gamma + 1} \, ,
\label{eq:8.1_sarazin_shock}
\end{equation}
where $P_1$ and $P_2$ are the pre- and post-shock pressure, and
$\gamma = 5/3$ is the adiabatic index of the gas.
For bow shocks in front of cold front, the shock may be conical at the
Mach angle, $\theta_m = \csc^{-1} ( {\cal M} )$.
The ratio of the pressure at the stagnation point in front of a cold front
to the pressure well ahead of it is given by
\begin{equation}
\frac{P_{\rm st}}{P_1} = \left\{
\begin{array}{cl}
\scriptstyle
\left( 1 + \frac{\gamma_ - 1}{2} {\cal M}^2
\right)^{\frac{\gamma}{\gamma - 1}} \, , &
\scriptstyle
{\cal M} \le 1 \, , \\
\scriptstyle
{\cal M}^2 \,
\left( \frac{\gamma + 1}{2}
\right)^{\frac{\gamma + 1}{\gamma - 1}} \,
\left( \gamma - \frac{\gamma - 1}{2 {\cal M}^2} \,
\right)^{- \frac{1}{\gamma - 1}} \, , &
\scriptstyle
{\cal M} > 1 \, . \\
\end{array}
\right.
\label{eq:8.1_sarazin_stagnation}
\end{equation}
If the motion of the cold front is supersonic, the stand-off distance
between the cold front and the bow shock varies inversely with $\cal M$.
For major mergers, these kinematic diagnostics generally indicate that
mergers are mildly transonic ${\cal M} \approx 2$, corresponding to
merger velocities of $\sim$2000 km/s.

Mergers also provide a useful environment for testing the role of various
physical processes in clusters.
For example, the very steep temperature gradients at cold fronts imply that
thermal conduction is suppressed by a large factor ($\sim$$10^2$;
Ettori \& Fabian 2000;
Vikhlinin, Markevitch, \& Murray 2001a),
presumably by magnetic fields.
The smooth front surfaces of some cold fronts suggest that Kelvin-Helmholtz
instabilities are being suppressed, also probably by magnetic fields.
Recently, Markevitch et al.\ (2004) have used the relative distributions of
dark matter, galaxies, and gas in the dramatic merging cluster
1E0657-56 (Figure~\ref{fig:8.1_sarazin_1e})
to argue that the collision cross-section per unit
mass of the dark matter must be low, $\sigma/m \la 1$ cm$^2$/g, which
excludes most of the self-interacting dark matter models invoked to explain
the mass profiles of galaxies.

\begin{figure}
\begin{center}
\includegraphics[width=\columnwidth]{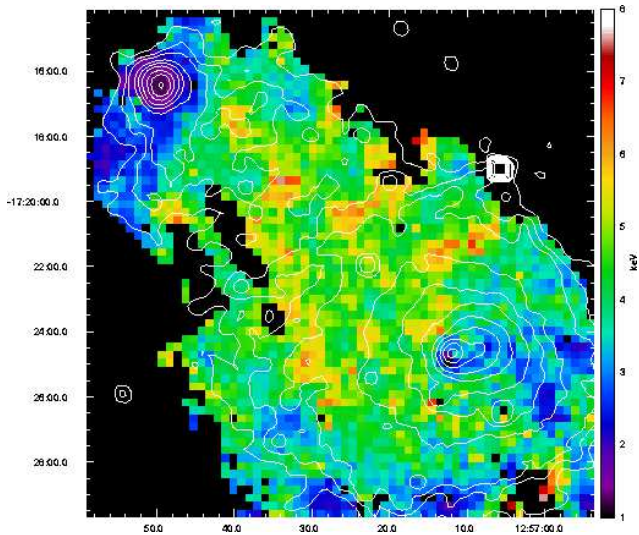}
\caption{\small
An X-ray temperature map (colors) of the merging cluster
Abell 1644 from the XMM/Newton observation of
Reiprich et al.\ (2004).
The contours show the X-ray surface brightness.
A trail of cooling gas lies below the subcluster at the
northeast corner of the image.}
    \label{fig:8.1_sarazin_a1644}
  \end{center}
\end{figure}

\begin{figure}
\begin{center}
\includegraphics[width=\columnwidth]{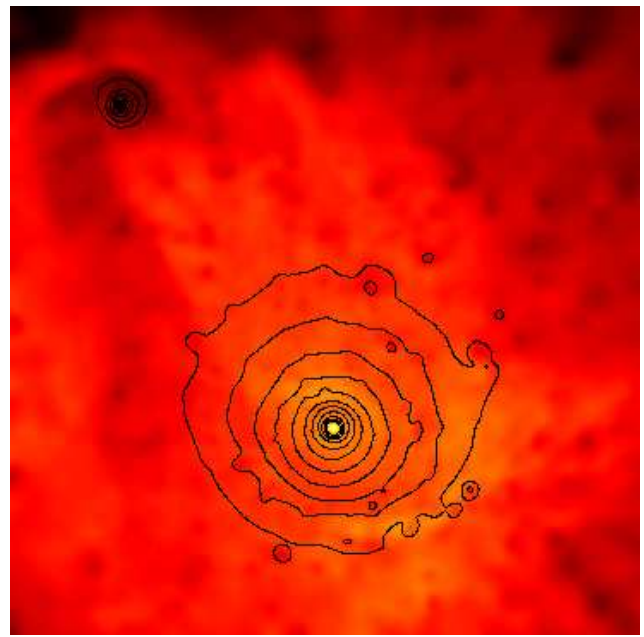}
\caption{\small
A temperature map (yellow hot, dark cold) from a hydrodynamical
simulation of a merging cluster
(Reiprich et al.\ 2004),
showing a cool trail behind the merging cluster.
The contours show the X-ray surface brightness.
Although the simulation was constructed independently of the observation
of Abell 1644 (Figure~\ref{fig:8.1_sarazin_a1644}),
the temperature map and X-ray image are similar.}
    \label{fig:8.1_sarazin_a1644sim}
  \end{center}
\end{figure}

In addition to merger cold fronts, cool trails of X-ray gas have been seen
behind merging subclusters in some Chandra and XMM/Newton observations.
For example,
Figure~\ref{fig:8.1_sarazin_a1644} shows the XMM/Newton X-ray temperature
map of the merging cluster Abell 1644
(Reiprich et al.\ 2004).
A trail of cooler gas lies below (behind) the merging subcluster.
Presumably, this gas was stripped from the subcluster during the merger.
Figure~\ref{fig:8.1_sarazin_a1644sim} shows the results of a numerical
hydrodynamical simulation of a cluster merger by Eric Tittley
(Reiprich et al.\ 2004).
The simulation shows a cool trail very similar to that in Abell 1644.

\section{Non-thermal effects of mergers}
\label{sec:8.1_sarazin_nonthermal}

High speed astrophysical shocks in diffuse gas generally lead to
significant acceleration of relativistic electrons.
For example, typical supernova remnants have blast wave shock velocities of
a few thousand km/s, which are comparable to the speeds in merger shocks.
(However, the Mach numbers in merger shocks are much lower.)
The ubiquity of radio emission from Galactic supernova remnants implies
that at least a few percent of the shock energy goes into accelerating
relativistic electrons, with more probably going into ions.
If these numbers are applied to strong merger shocks in clusters, one would
expect that relativistic electrons with a total energy of
$E_{\rm rel,e} \sim 10^{62}$ erg would be accelerated, with even more
energy
in the relativistic ions.
Thus, merging clusters should have huge populations of relativistic
particles.

\begin{figure}
\begin{center}
\includegraphics{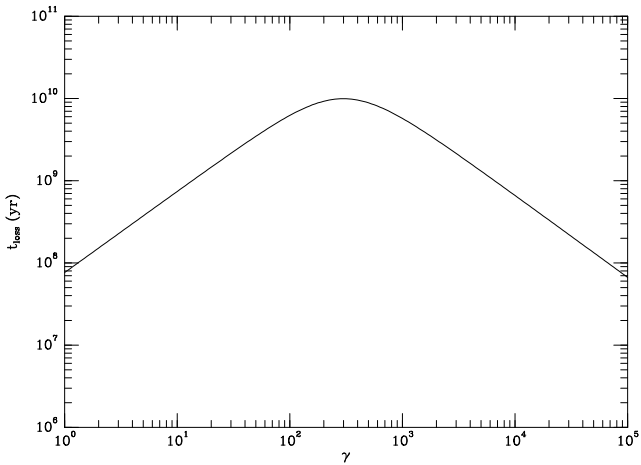}
\vskip2.5truein
\caption{\small
The lifetimes of relativistic electrons in a typical cluster as
a function of their Lorentz factor $\gamma$
(Sarazin 1999a).
The lifetime is maximum for $\gamma \sim 300$.}
    \label{fig:8.1_sarazin_lifetime}
  \end{center}
\end{figure}

Clusters may contain both primary and secondary relativistic electrons.
Primary electrons are accelerated directly in mergers.
Secondary electron are produced by the interactions of relativistic
ions with thermal ions in the intracluster medium.
Collisions between relativistic ions (mainly protons) and thermal
ions (also mainly protons) can produce pions (and other mesons):
\begin{equation}
p + p \rightarrow p + p + n \pi \, .
\label{eq:8.1_sarazin_secondary1}
\end{equation}
The charged pions decay to produce electrons and positrons:
\begin{eqnarray}
\pi^{\pm} & \rightarrow & \mu^{\pm} + \nu_{\mu} ( \bar{\nu}_{\mu} )
\nonumber \\
\mu^{\pm} & \rightarrow & e^{\pm} + \bar{\nu}_{\mu} ( \nu_{\mu} ) +
\nu_{e} ( \bar{\nu}_{e} ) \, .
\label{eq:8.1_sarazin_secondary2}
\end{eqnarray}

In mergers, primary relativistic particles may be produced 
through shock acceleration or by turbulent acceleration,
where the turbulence may have been generated by a merger shock passage.
Often, it is argued that cluster radio relics are due to
shock acceleration, while cluster radio halos are produced by
turbulent acceleration.
Although the seed particles for acceleration could come from the
thermal intracluster gas, it is easier to re-accelerate a relic
population of low energy relativistic particles.

However they are produced,
clusters should retain some of these relativistic particles for very long times.
The cosmic rays gyrate around magnetic field lines, which are frozen-in to
the gas, which is held in by the strong gravitational fields of clusters.
Because clusters are large, the timescales for diffusion are generally
longer than the Hubble time.
The low gas and radiation densities in the intracluster medium imply that
losses by relativistic ions are very slow, and those by relativistic
electrons are fairly slow.

Figure~\ref{fig:8.1_sarazin_lifetime}
shows the loss timescale for electrons under typical cluster
conditions;
electrons with Lorentz factors $\gamma \approx 300$ and energies of
$\approx$150 MeV have lifetimes which approach the Hubble time, as long
as cluster magnetic fields are not too large ($B \la 3$ $\mu$G).
On the other hand, the higher energy particles which produce radio
synchrotron emission and could produce hard X-ray emission have
relatively short lifetimes, comparable to the durations of cluster
mergers.
As a result, clusters should contain two populations of primary
relativistic
electrons accelerated in mergers:
those at $\gamma \sim 300$ which have been produced by mergers
over the lifetime of the clusters; and a tail to higher energies produced
by any current merger.

\begin{figure}
\begin{center}
\includegraphics{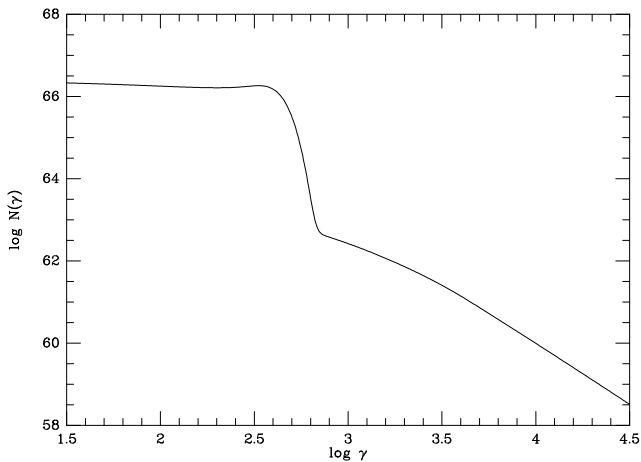}
\vskip2.5truein
\caption{\small
The energy spectrum of relativistic electrons in a model for
a merging cluster
(Sarazin 1999a).
The large population at $\gamma \sim 300$ are due to many previous mergers,
while the tail to high energies is due to the current merger.}
    \label{fig:8.1_sarazin_espect}
  \end{center}
\end{figure}

Figure~\ref{fig:8.1_sarazin_espect} shows the energy spectrum of primary
electrons in an example of a model for a cluster undergoing a merger
(Sarazin 1999a).
There is a large population of low energy electrons ($\gamma \sim 300$,
$ E \sim  150$ MeV), which were accelerated in earlier mergers in the
same cluster.
There is also an approximately power-law tail of higher particles in
the  electron distribution;
these electrons are being accelerated in the current merger.

\begin{figure}
\begin{center}
\includegraphics{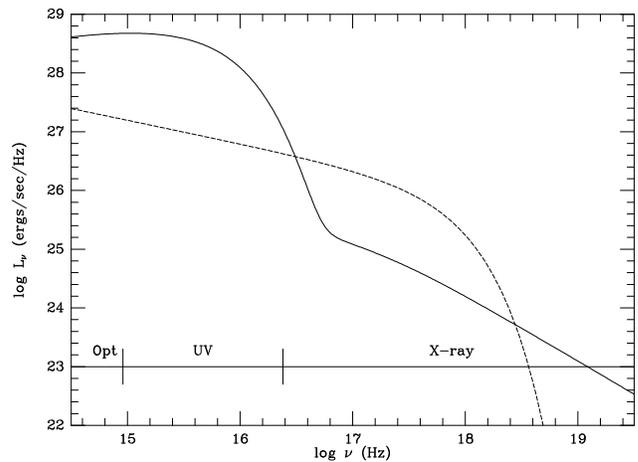}
\vskip2.5truein
\caption{\small
The IC spectrum produced by the electrons in the model for a merging
cluster in
Figure~\ref{fig:8.1_sarazin_espect}.
For reference, the dashed line is thermal bremsstrahlung at a typical
cluster luminosity.}
    \label{fig:8.1_sarazin_ic}
  \end{center}
\end{figure}

The IC emission spectrum from the electrons in this model is
shown in Figure~\ref{fig:8.1_sarazin_ic}.
The lower energy electrons ($\gamma \sim 300$) will mainly be visible in
the EUV/soft X-ray range
(e.g., Sarazin \& Lieu 1998).
Because these low energy electrons have very long lifetimes, they should
be present in most clusters.
Such EUV/soft X-ray emission has been apparently been seen in several
clusters
(e.g., Nevalainen et al.\ 2003;
Bowyer et al.\ 2004),
although its origin is uncertain and may well be thermal
(e.g., Kaastra et al.\ 2003).

More energetic electrons, with energies of many GeV, produce hard X-ray
IC emission and radio synchrotron emission.
Diffuse radio sources, not associated with radio galaxies, have been
observed for many years in merging clusters
(see the review by Feretti in this volume).
Centrally located, unpolarized, regular sources are called ``radio halos'',
while peripheral, irregular, polarized sources are called ``radio
relics''.
Radio halos and relics are only found in clusters which are undergoing
mergers.
Recent Chandra observations seem to show a direct connection between
radio halos and merger shocks in clusters
(Markevitch \& Vikhlinin 2001;
Govoni et al.\ 2004; see also the papers by Feretti and by Markevitch
in this volume).

The radio power in cluster radio halos correlates very strongly with
the X-ray luminosity and X-ray temperature of clusters
(e.g., Bacchi et al 2003;
Liang et al.\ 2000).
The correlation is much stronger than expected based on a
simple scaling of cluster properties
(e.g., Kempner \& Sarazin 2001).
This strong correlation might be explained if the radio-emitting
electrons are due to merger shock or turbulent acceleration, and
if the high X-ray luminosities and temperatures
are due to the transient boosts which occur during mergers
(Randall et al.\ 2002;
Randall \& Sarazin 2004).

The same higher energy electrons responsible for the radio emission
will produce hard X-ray IC emission.
Recently, such emission appears to have been detected with BeppoSAX
and RXTE in at least Coma
(Fusco-Femiano et al.\ 1999;
Rephaeli, Gruber, \& Blanco 1999),
Abell 2256
(Fusco-Femiano et al.\ 2000;
Rephaeli \& Gruber 2003),
and
Abell 754
(Fusco-Femiano et al.\ 2003).
However, the detections are relatively weak and controversial
(Fusco-Femiano et al.\ 2004;
Rossetti \& Molendi 2004).
Observations with INTEGRAL may help to resolve the nature of the hard
excesses in clusters by providing higher spatial resolution
hard X-ray images.

One of the difficulties in detecting nonthermal hard X-ray excesses in clusters
of galaxies is the very strong thermal emission from clusters.
As noted above, clusters with radio halos tend to be particularly hot
(Liang et al.\ 2000),
which only adds to this difficulty.
One possible way around this difficulty would be to find
cooler groups with radio halos;
then, the contrast of the hard X-ray IC emission with the group thermal
emission would be more favorable.
A survey for radio halos in groups and follow-up hard X-ray observations
might be useful.
It is possible that one case may have been detected already in IC 1262
(Hudson \& Henriksen 2003).

\begin{figure}
\begin{center}
\includegraphics{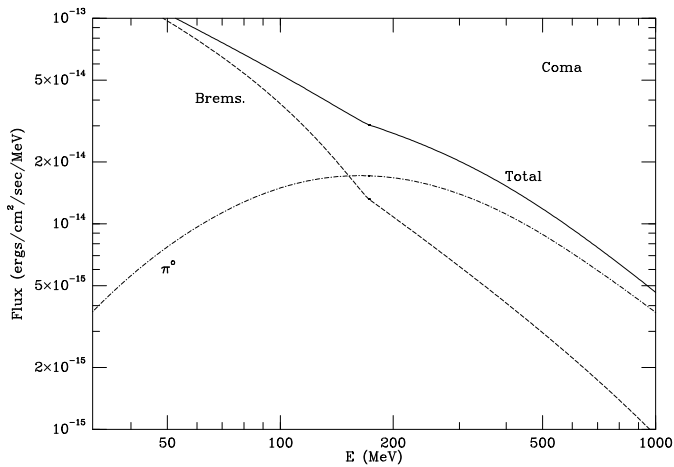}
\vskip2.5truein
\caption{\small
The gamma-ray emission spectrum from a model for relativistic particles in
the Coma cluster
(Sarazin 1999b).
The emission from electrons (mainly bremsstrahlung in this model)
and ions (due to $\pi^o$ decays) are shown separately.}
    \label{fig:8.1_sarazin_gamma}
  \end{center}
\end{figure}

I believe one of the most exciting possibilities for the future is the
detection of clusters in hard gamma-ray radiation
(Figure~\ref{fig:8.1_sarazin_gamma}).
Essentially, all models for the nonthermal populations in clusters predict
that they should be very luminous gamma-ray sources, particularly at photon
energies of $\sim$100 MeV
(Sarazin 1999b;
Gabici \& Blasi 2004).
The emission at these energies is partly due to electrons with energies
of $\sim$150 MeV, which should be ubiquitous in clusters.
One nice feature of this spectral region is that emission is produced both
by relativistic electrons (through bremsstrahlung and IC emission)
and relativistic ions.
The ions (mainly protons) produce gamma-rays by $\pi^o$ decay;
the process is similar to the secondary electron production process
in
Eq.\ (\ref{eq:8.1_sarazin_secondary1})
\&
(\ref{eq:8.1_sarazin_secondary2}):
\begin{eqnarray}
p + p & \rightarrow & p + p + n \pi \nonumber \\
\pi^o & \rightarrow & 2 \gamma \, .
\label{eq:8.1_sarazin_gammapi}
\end{eqnarray}
Both the emissivity of bremsstrahlung by relativistic electrons and
that of gamma-ray emission by $\pi^o$ decay involve collisions
with the ions in the thermal intracluster gas;
thus, the ratio of the two emission mechanisms depends mainly on
the ratio of relativistic electrons to relativistic ions.
Since the two emission mechanisms have different spectral shapes and are
expected to be of comparable importance,
one can determine separately both population is clusters.
Models suggest that GLAST and AGILE will detect $\ga$40 nearby clusters
(e.g., Gabici \& Blasi 2004).

\section*{Acknowledgments}
I would like to thank my collaborators, who have included
Liz Blanton,
Tracy Clarke,
Josh Kempner,
Scott Randall,
Thomas Reiprich,
Paul Ricker,
and
Eric Tittley.
I thank Maxim Markevitch for providing Figure~\ref{fig:8.1_sarazin_1e}.
This work was supported by the National
Aeronautics and Space Administration through
Chandra Award Numbers
GO2-3159X,
GO3-4155X,
GO3-4160X,
GO4-5149X,
and
GO4-5150X,
and
XMM/Newton Award Numbers
NAG5-13737
and
NAG5-13088.


\begin{thebibliography}{}
\setlength\itemsep{0cm}

\bibitem[xxx(1950)]{zzz}
Bacchi, M., Feretti, L., Giovannini, G., \& Govoni, F. 2003, A\&A, 400, 465

\bibitem[xxx(1950)]{zz}
Bowyer, S., Korpela, E. J., Lampton, M., Jones, T. W. 2004, ApJ, 605, 168

\bibitem[xxx(1950)]{}
Buote, D. A., \& Tsai, J. C. 1996, ApJ, 458, 27

\bibitem[xxx(1950)]{a}
Ettori, S., \& Fabian, A. C. 2000, MNRAS, 317, L57

\bibitem[Fusco-Femiano et al.(1999)]{FFea1}
Fusco-Femiano, R., et al., 1999, ApJ, 513, L21

\bibitem[Fusco-Femiano et al.(2000)]{FFea2}
Fusco-Femiano, R., et al., 2000, ApJ, 534, L7

\bibitem[Fusco-Femiano et al.(2000)]{FFea3}
Fusco-Femiano, R., et al., 2003, A\&A, 398, 441

\bibitem[xxx(1950)]{aa}
Fusco-Femiano, R., Orlandini, M., Brunetti, G., Feretti,
L., Giovannini, G., Grandi, P., \& Setti, G.
2004, ApJ, 602, L73

\bibitem[xxx(1950)]{s}
Gabici, S., \& Blasi, P.  2004, APh, 20, 579

\bibitem[xxx(1950)]{ssss}
Govoni, F., Markevitch, M., Vikhlinin, A., VanSpeybroeck, L., Feretti, L.,
\& Giovannini, G. 2004, ApJ, 605, 695

\bibitem[xxx(1950)]{sss}
Hudson, D. S., \& Henriksen, M. J. 2003, ApJ, 595, L1

\bibitem[xxx(1950)]{ss}
Kaastra, J. S., Lieu, R., Tamura, T., Paerels, F. B. S., \& den Herder, J. W.
2003, A\&A, 397, 445

\bibitem[Kempner \& Sarazin(2001)]{KS2}
Kempner, J., \& Sarazin, C. L. 2001, ApJ, 548, 639

\bibitem[xxx(1950)]{d}
Kempner, J., Sarazin, C. L., \& Ricker, P. R. 2002, ApJ, 579, 236

\bibitem[Liang et al.(2000)]{LHBA}
Liang, H., Hunstead, R. W., Birkinshaw, M., \& Andreani, P. 2000, ApJ,
544, 686

\bibitem[xxx(1950)]{f}
Markevitch, M., et al., 2000, ApJ, 541, 542

\bibitem[xxx(1950)]{g}
Markevitch, M., et al.,
2004, ApJ, in press (astro-ph/0309303)

\bibitem[xxx(1950)]{h}
Markevitch, M., \& Vikhlinin, A. 2001, ApJ, 563, 95

\bibitem[xxx(1950)]{hh}
Nevalainen, J., Lieu, R., Bonamente, M., \& Lumb, D. 2003, ApJ, 584, 716

\bibitem[xxx(1950)]{j}
Randall, S. W., \& Sarazin, C. L. 2004, preprint

\bibitem[xxx(1950)]{i}
Randall, S. W., Sarazin, C. L., \& Ricker, P. M. 2002, ApJ, 577, 579

\bibitem[xxx(1950)]{ii}
Randall, S. W., Sarazin, C. L., \& Ricker, P. M. 2004, preprint

\bibitem[xxx(1950)]{jj}
Reiprich, T. H., Sarazin, C. L., Kempner, J. C., \&
Tittley, E. 2004, ApJ, in press (astro-ph/0308282)

\bibitem[Rephaeli \& Gruber(2003)]{RG}
Rephaeli, Y., \& Gruber, D. 2003, ApJ, 579, 587

\bibitem[Rephaeli \& Gruber(1999)]{RGB}
Rephaeli, Y., Gruber, D., \& Blanco, P. 1999, ApJ, 511, L21

\bibitem[xxx(1950)]{k}
Ricker, P. M., \& Sarazin, C. L. 2001, ApJ, 561, 621

\bibitem[xxx(1950)]{l}
Rossetti, M., \& Molendi, S. 2004, A\&A, 414, L41

\bibitem[xxx(1950)]{q}
Sarazin, C. L. 1999a, ApJ, 520, 529

\bibitem[xxx(1950)]{w}
Sarazin, C. L. 1999b, in Diffuse Thermal and Relativistic Plasma in Galaxy
Clusters, ed.\ H. B\"ohringer, L. Feretti, \& P. Schuecker (Garching: MPE
Rep.\ 271), 185

\bibitem[xxx(1950)]{e}
Sarazin, C. L. 2002, in Merging Processes in Clusters of Galaxies,
ed. L. Feretti, I. M. Gioia, \& G. Giovannini (Dordrecht: Kluwer), 1

\bibitem[xxx(1950)]{ee}
Sarazin, C. L., \& Lieu, R. 1998, ApJ, 494, L177

\bibitem[xxx(1950)]{r}
Smith, G. P., Edge, A. C., Eke, V. R., Nichol, R. C.,
Smail, I., \& Kneib, J.-P. 2003, ApJ, 590, L79

\bibitem[xxx(1950)]{t}
Torri, E., Meneghetti, M., Bartelmann, M., Moscardini, L., Rasia, E., \&
Tormen, G. 2004, MNRAS, 349, 476

\bibitem[xxx(1950)]{y}
Vikhlinin, A., Markevitch, M., \& Murray, S. M. 2001a, ApJ, 549, L47

\bibitem[xxx(1950)]{u}
Vikhlinin, A., Markevitch, M., \& Murray, S. M. 2001b, ApJ, 551, 160

%
%
%
%
%

\end{thebibliography}
\end{document}